\documentstyle[aps,pre,preprint]{revtex} 

\newcommand{\R}{\hbox{$\hbox{I}\mskip-4mu\hbox{R}$}}
\tightenlines

\begin{document}

\draft
\title{Test your surrogate data before you test for nonlinearity}
\author{D. Kugiumtzis \thanks{E-mail: dimitris@mpipks-dresden.mpg.de}}
\address{Max-Planck Institute for Physics of Complex Systems\\
N{\"o}thnitzer Str. 38, 01187 Dresden, Germany}
\date{\today}
\maketitle

\begin{abstract}
The schemes for the generation of surrogate data in order to test the null
hypothesis of linear stochastic process undergoing nonlinear static 
transform are investigated as to their consistency in representing the 
null hypothesis. 
In particular, we pinpoint some important caveats of the prominent algorithm
of amplitude adjusted Fourier transform surrogates (AAFT) and compare it to
the iterated AAFT (IAAFT), which is more consistent in representing the null
hypothesis.    
It turns out that in many applications with real data the inferences of
nonlinearity after marginal rejection of the null hypothesis were premature
and have to be re-investigated taken into account the inaccuracies in the AAFT
algorithm, mainly concerning the mismatching of the linear correlations.   
In order to deal with such inaccuracies we propose the use of linear together
with nonlinear polynomials as discriminating statistics.  
The application of this setup to some well-known real data sets cautions
against the use of the AAFT algorithm.   
\end{abstract}


% ======================
\section{Introduction}
% ======================

Often an indirect approach is followed to investigate the existence of
nonlinear dynamics in time series by means of hypothesis testing using
surrogate data \cite{Theiler92,Theiler92b}.   
To this respect, the null hypothesis of a linear stochastic process undergoing
a nonlinear static transform is considered as the most appropriate
because it is the closest to nonlinearity one can get with linear dynamics.   
Surrogate data representing this null hypothesis ought to be random data,
but possess the same power spectrum (or autocorrelation) and amplitude
distribution as the original time series. 
To test the null hypothesis, a method sensitive to nonlinearity is applied to
the original time series and to a set of surrogate time series. 
The null hypothesis is rejected if a statistic derived from the method 
statistically discriminates the original from the surrogate data. 

For the generation of surrogate data the algorithm of the so-called amplitude
adjusted Fourier transform (AAFT), by Theiler and co-workers
\cite{Theiler92,Theiler92b}, has been followed in a number of applications so
far
\cite{Prichard93,Rombouts95,Yip95,Pritchard95,Pradhan96,Ivanov96,Theiler96,Menendez97,Stam97,Govindan98}.

Recently, another algorithm similar to that of Theiler, but making 
use of an iterative scheme in order to achieve arbitrarily close approximation
to the autocorrelation and the amplitude distribution was proposed by
Schreiber and Schmitz \cite{Schreiber96}. 
We refer to it as the iterative AAFT (IAAFT) algorithm.
A more advanced algorithm designed to generate surrogates for any given
constraints is supposed to be very accurate for moderate and large data sets,
but in the deterrent cost of long computation time \cite{Schreiber98}.  
Therefore it is not considered in our comparative study.

Other schemes circumvent the problem of non-Gaussian distribution
by making first a so-called ''Gaussianization'' of the original data
and proceed with this data set generating Fourier transform (FT) surrogates
instead \cite{Palus95,Muller96,Casdagli97}.  
In this way the results of the FT surrogate test concern the ''Gaussianized''
data and it is not clear why the same results should be valid for the original
non-Gaussian data.

Shortcomings of the AAFT algorithm due to the use of FT on short periodic
signals and signals with long coherent times have been reported elsewhere
\cite{Theiler93,Palus95,Stam98}.   
Here, we pinpoint more general caveats of the method, often expected to occur
in applications, and give comparative results with the IAAFT method.  
  
In Section\ \ref{Algorithms}, the AAFT and IAAFT algorithms are presented and
discussed. In Section\ \ref{Bias}, the dependence of the hypothesis test on the
generating scheme for the surrogate data is examined and in
Section\ \ref{Applications} the performance of the algorithms on real data is
presented.  

% =====================================
\section{The AAFT and IAAFT algorithms}
\label{Algorithms}
% =====================================

Let $\{ x_i \}, \, i\!=\!1,\ldots,N$, be the observed time series. 
According to the null hypothesis, $x_i \!=\! h(s_i)$, where $\{ s_i \}, \,
i\!=\!1,\ldots,N$, is a realization of a Gaussian stochastic process (and thus
linear) and $h$ is a static measurement function, possibly nonlinear. 
In order for a surrogate data set $\{ z \}$ (of the same length $N$) to
represent the null hypothesis it must fulfill the following two conditions:
1) $R_z(\tau) \!=\! R_x(\tau)$ for $\tau \!=\! 1,\ldots,\tau^{\prime}$, 
2) $A_z(z) \!=\! A_x(x)$, where $R$ is the autocorrelation, $A$ the empirical
amplitude distribution and $\tau^{\prime}$ a sufficiently large lag time. 

% -----------------------------
\subsection{The AAFT algorithm}
% -----------------------------
  
The AAFT algorithm starts with the important assumption that $h$ is a monotonic
function, i.e., $h^{-1}$ exists. 
The idea is to simulate first $h^{-1}$ (by
reordering white noise data to have the same rank structure as $\{ x \}$, call
it $\{ y \}$, step\ 1), then make a randomized copy of the obtained version of
$\{ s \}$ (by making a FT surrogate of $\{ y \}$, call it $\{ y^{\text{FT}}
\}$, step\ 2) and transform it back simulating $h$ (by reordering $\{ x \}$
to have the same rank structure as $\{ y^{\text{FT}} \}$, call it $\{ z \}$,
step\ 3).   

In step\ 1, it is attempted to bring $\{ x \}$ back to $\{ s \}$, in a loose
statistic manner, by constructing a time series with similar structure to $\{ x
\}$, but with Gaussian amplitude distribution. 
However, it is not clear what is the impact of this process on the
autocorrelation $R$.  
From the probabilistic theory for transformations of stochastic variables it
is known that $R_x \!\leq\! |R_s|$ in general \cite{Johnson70}, and moreover
$R_x \!=\! g(R_s)$, for a smooth function $g$ (under some assumptions on $h$).
Assuming that $h^{-1}$ exists and is successfully simulated in step\ 1, we get
$R_y \!\approx\! R_s$ and thus $R_x \!\leq\! |R_y|$.
Analytic verification of this is not possible because reordering constitutes
an ''empirical transformation''.
When $h$ is not monotonic or not successfully simulated by the reordering in
step\ 1 the relation $R_y \! \approx\! R_s$ cannot be established and $R_y$
will be somehow close to $R_x$.

The phase randomization process in step\ 2 does not affect $R$ apart from
statistical fluctuations ($R_y \!\approx\! R_y^{\text{FT}}$), neither alter
the Gaussian distribution, but it just destroys any possible nonlinear
structure in $\{ y \}$.  
The reordering in step\ 3 gives $A_z(z) \!=\! A_x(x)$.  
This process changes also $R$ according to the function $g$, i.e., $R_z \!=\!
g(R_y)$, assuming again that $h$ is successfully simulated by the reordering.
For the latter to be true, a necessary condition is $R_z \!\leq\! |R_y|$.
So, the preservation of the autocorrelation $R_z \!\approx\! R_x$ is
established only if in step\ 1 $R_y \!\approx\! R_s$ is achieved after the
reordering, which is not the case when $h$ is not monotonic.
Otherwise, the AAFT algorithm gives biased autocorrelation with a bias
determined by the reordering in step\ 1 and the form of $g$, and subsequently
the form of $h$.

To elucidate, we consider the simplest case of an AR($1$) process [$s_i \!=\! b
s_{i-1} \!+\! w_i$, $b\!=\!0.4$ and $w_i \!\sim\! {\cal N}(0,1-b^2)$], 
and static power transforms, $x \!=\! h(s) \!=\! s^a$, for positive $a$. 
For $s_i \!\in\! \R$, $h^{-1}$ exists only for odd values of $a$. 
For even values of $a$, a deviation of $R_y$ from $R_s$ is expected resulting
to surrogate data with different autocorrelation than that of the original
($R_z \!\neq\! R_x$).   
Monte Carlo simulation approves this finding as shown in Fig.\
\ref{AR1Autoc}(a) ($N\!=\!2048$, $a\!=\!1,2,\ldots,10$, $M\!=\!40$ surrogates,
$100$ realizations). 
% ----------------------
%           Fig.1
% ----------------------
Note that for $R_y(1)$ the standard deviation (SD) is almost zero indicating
that all $40$ reordered noise data $\{ y \}$ obtain about the same
$R_y(1)$ for each realization of $\{ s \}$, which depends on $R_s(1)$ at each
realization.  
The results show the good matching $R_y(1) \!\approx\! R_s(1)$ and $R_z(1) \!
\approx\! R_x(1)$ for odd $a$.
For even $a$, $R_y(1)$ is always on the same level, well below $R_s(1)$,
and $R_z(1) \!<\! R_x(1)$, with the difference to decrease with larger powers.

% ------------------------------
\subsection{The IAAFT algorithm}
\label{IAAFT}
% ------------------------------

The IAAFT algorithm makes no assumption for the form of the $h$ transform
\cite{Schreiber96}. 
Starting with a random permutation $\{ z^{(0)} \}$ of the data $\{ x \}$, the
idea is to repeat a two-step process: approach $R_x$ in the first step (by
bringing the power spectrum of $\{ z^{(i)} \}$ to be identical to that of $\{
x \}$, call the resulting time series $\{ y^{(i)} \}$), and regain the
identical $A_x$ in the second step (by reordering $\{ x \}$ to have the same
rank structure as $\{ y^{(i)} \}$, call it $\{ z^{(i+1)} \}$). 

The desired power spectrum gained in step\ 1 is changed in step\ 2 and
therefore several iterations are required to achieve convergence of the power
spectrum.
The algorithm terminates when the exact reordering is repeated in two
consecutive iterations indicating that the power spectrum of the surrogate
cannot be brought closer to the original.
The original algorithm gives as outcome the data set derived from step\ 2 at
the last iteration ($\{ z^{(i+1)} \}$, where $i$ the last iteration),
denoted here IAAFT-1.
The IAAFT-1 surrogate has exactly the same amplitude distribution as the
original, but discrepancies in autocorrelation are likely to occur. 
If one seeks the best matching in autocorrelation leaving the
discrepancies for the amplitude distribution, the data set from step\ 1 after
the last iteration ($\{ y^{(i)} \}$), denoted IAAFT-2, must be chosen instead. 

The bias of the autocorrelation with IAAFT-1 can be significant, as shown
in Fig.\ \ref{AR1Autoc}(b). 
For reasons we cannot explain, the bias in $R(1)$ gets larger for odd values of
$a$ (monotonic $h$), for which AAFT gives no bias (the same bias was observed
using 16384 samples ruling out that it can be due to limited data size).  
On the other hand, the matching in autocorrelation with IAAFT-2 was perfect
(not shown). 
The surrogate data derived from the IAAFT algorithm exhibit little (IAAFT-1)
or essentially zero (IAAFT-2) variation in autocorrelation compared to AAFT. 
This is an important property of the IAAFT algorithm in general because it
increases the significance of the discriminating statistic as it will be shown
below.  

% ==========================================================
\section{The effect of biased autocorrelation}
\label{Bias}
% ==========================================================

By construction, the IAAFT algorithm can represent the null hypothesis
regardless of the form of $h$ while the AAFT algorithm cannot represent it
when $h$ is not monotonic. 
One can argue that the deviation in the autocorrelation due to possible
non-monotonicity of $h$ is small and does not affect the results of the
test.  
This is correct only for discriminating statistics which are not sensitive to
linear data correlations, but most of the nonlinear methods, including all
nonlinear prediction models, are sensitive to data correlations and therefore
they are supposed to have the power to distinguish nonlinear correlations from
linear.   

We show in Fig.\ \ref{AR1Rej} the results of the test with AAFT and IAAFT
surrogate data from the example with the AR(1) process. 
% ----------------------
%           Fig.2
% ----------------------
Two discriminating statistics are drawn from the correlation coefficient
of the one step ahead fit $\rho(1)$ using an AR(1) model and from the
$\rho(1)$ using a local average mapping (LAM) (embedding dimension $m\!=\!3$,
delay time $\tau\!=\!1$, neighbors $k\!=\!5$).  
The significance of each statistic $q$ [here $q \!=\! \rho(1)$] is computed
as 
\begin{equation}
  S = \frac{|q_0 - \bar{q}|}{\sigma_q}
\label{sigma}
\end{equation}
where $q_0$ is for the original data, $\bar{q}$ and $\sigma_q$ is the mean
and SD of $q_i$, $i\!=\!1,\ldots,M$, for the $M$ surrogate data (here
$M\!=\!40$). 
The significance $S$ is a dimensionless quantity, but it is customarily given
in units of ''sigmas'' $\sigma$.
A value of $2\sigma$ suggests the rejection of the null hypothesis at the
$95\%$ confidence level.

For AAFT, the number of rejections with both AR(1) and LAM is at the level of
the ''size'' of the test ($5\%$, denoted with a horizontal line in
Fig.\ \ref{AR1Rej}) when $h$ is monotonic, but at 
much higher levels when $h$ is not monotonic, depending actually on the
magnitude of $a$. 
For the IAAFT algorithm the results are very different.
Using IAAFT-1 surrogates the number of rejections is almost always much larger
than the ''size'' of the test and the opposite feature from that for AAFT is
somehow observed for the even and odd values of $a$. 
The extremely large number of rejections with IAAFT-1 is actually due to the
small variance of the statistics for the IAAFT surrogates [see also Fig.\
\ref{AR1Autoc}(b)].  
On the other hand, the $\rho(1)$ of the AR(1) for the IAAFT-2 surrogates seems
to coincide with the original because, besides the small SD in Eq.\
(\ref{sigma}), the significance is almost always below $2\sigma$.
Note that the $\rho(1)$ of AR(1) behaves similar to $R(1)$ for this particular
example.
The values of $\rho(1)$ of LAM for each surrogate data group are more spread 
giving thus less rejections for AAFT and IAAFT-1.
For IAAFT-2 and for $a\!=\!3$ and $a\!=\!5$, there are too many rejections and
this cannot be easily explained since this behavior is not observed for the
linear measure.  
We suspect that the rejections are due to the difference in the amplitude
distribution of the IAAFT-2 surrogate data from the original, which may effect
measures based on the local distribution of the data such as LAM.
  
Simulations with AR-processes of higher order and with stronger correlations
showed qualitatively the same results. 

% ========================================================
\section{Surrogate tests with AAFT and IAAFT on real data}
\label{Applications}
% ========================================================

In order to verify the effect of the bias in autocorrelation in a more
comprehensive manner, we consider in the following examples with real data the 
discriminating statistics $q^i \!=\! \rho(T,i)$, $i\!=\!1,\ldots,n$, from
the $T$ time step ahead fit with polynomials $p_i$ of the Volterra series of
degree $d$ and memory (or embedding dimension) $m$ 
\begin{eqnarray}
  \hat{x}_{i+T} & = & p_n(\bbox{x}_i) = p_n(x_i,x_{i-1},\ldots,x_{i-(m-1)})
                \nonumber \\ 
                & = & a_0 + a_1 x_i + \ldots + a_m x_{i-(m-1)} \\
                &   & + a_{m+1} x_i^2 + a_{m+2} x_i x_{i-1} + \ldots + 
                  a_{n-1} x_{i-(m-1)}^d \nonumber 
\end{eqnarray}
where $n \!=\! (m+d)! / (m!d!)$ is the total number of terms. 
To distinguish nonlinearity, $d\!=\!2$ is sufficient.
Starting from $p_1 \!=\! a_0$, we construct $n$ polynomials adding one term at
a time. 
Note that $p_2,\ldots,p_{m+1}$ are actually the linear AR models of order
$1,\ldots,m$, respectively, and the first nonlinear term enters in the
polynomial $p_{m+2}$. 
So, if the data contains dynamic nonlinearity, this can be diagnosed by an
increase of $\rho(T,i)$ when nonlinear terms are added in the polynomial
form. 
This constitutes a direct test for nonlinearity, independent of the surrogate
test, and its significance is determined by the increase of $\rho(T,i)$
for $i\!>\!m+1$.
Note that a smooth increase of $\rho(T,i)$ with the polynomial terms is
always expected because the fitting becomes better even if only noise is
''modeled'' in this way. 
In Ref. \cite{Barahona96}, where this technique was applied first, this is
avoided by punishing the addition of terms and the AIC criterion is used
instead of $\rho$.  
Here, we retain the $\rho$ measure bearing this remark in mind when we
interpret the results from the fit.    

We choose this approach because on the one hand, it gives a clear indication
for the existence or not of nonlinearity, and on the other hand, it preserves
or even amplifies the discrepancies in the autocorrelation, so that we can
easily verify the performance of the AAFT and IAAFT methods. 

% ------------------------------
\subsection{The sunspot numbers} 
% ------------------------------
 
We start with the celebrated annual sunspot numbers (e.g. see \cite{Tong90}).  
Many suggest that this time series involves nonlinear dynamics
(e.g. see \cite{Tong90} for a comparison of classical statistical models on
sunspot data, and \cite{Weigend90,Kugiumtzis97} for a comparison of
empirical nonlinear prediction models).
The sunspot data follows rather a squared Gaussian than a Gaussian
distribution and therefore the AAFT does not reach a high level of accuracy in
autocorrelation [see Fig.\ \ref{SunspotsAutoc}(a)].   
% ----------------------
%           Fig.3
% ----------------------
Note that the $R_y(\tau)$ of the reordered noise data $\{ y \}$ follows well
with the $R_z(\tau)$ of the AAFT, and is even closer to the original
$R_x(\tau)$.  
This behavior is reminiscent to that of squared transformed AR-data [see
Fig.\ \ref{AR1Autoc}(a) for $a=2$], which is in agreement
with the comment in \cite{Schreiber98b} that the sunspot numbers
are in first approximation proportional to the squared magnetic field
strength.
The condition $R_z(\tau) \! \leq \! |R_y(\tau)|$ holds supporting that the
simulation of $h$ transform (step\ 3 of AAFT algorithm) is successful. 
Due to the short size of the sunspot data also IAAFT-1 surrogates cannot mimic
perfectly the autocorrelation, as shown in Fig.\ \ref{SunspotsAutoc}(b).
On the other hand, the IAAFT-2 surrogates match perfectly the autocorrelation
and follow closely the original amplitude distribution.

The discrepancies in autocorrelation are well reflected in the correlation
coefficient $\rho$ from the polynomial modeling as shown in Fig.\
\ref{SunspotsCC}. 
% ----------------------
%           Fig.4
% ----------------------
To avoid bias in favor of rejecting the null hypothesis, we use arbitrarily
$m\!=\!10$ in all our examples in this section. 
The fit of the original sunspot data is improved as linear terms increase from
$1$ to $9$, and no improvement is observed adding the tenth linear term which
is in agreement with the choice of AR(9) as the best linear model
\cite{Tong90}.  
As expected, this feature is observed for the AAFT and IAAFT surrogates as
well [see Fig.\ \ref{SunspotsCC}(a)].
Further, the inclusion of the first nonlinear term ($x_i^2$), improves the
fitting of the original sunspot numbers, but not of the surrogates.
Actually, the Volterra polynomial fitting shows that the interaction of $x_i$
and $x_{i-1}$ with theirselves and with each other is prevalent for the
sunspot data [note the increase of $\rho(1,i)$ for $i\!=\!12$, $i\!=\!13$, and
$i\!=\!22$].   
When compared with AAFT surrogates this feature is not so obvious mainly due
to the large variance of $\rho(1,i)$ of the AAFT surrogates and the large
discrepancy from the $\rho(1,i)$ of the original data, which persists also for
the linear terms (for $i\!=\!2,\ldots,11$).  
For the IAAFT-1 surrogate data, there is also a small discrepancy due to
imperfect matching of the autocorrelation, which disappears when the IAAFT-2
surrogate data are used instead.
  
The significance $S$ of the discriminating statistics $\rho(1,i)$,
$i\!=\!1,\ldots,66$, shown in Fig.\ \ref{SunspotsCC}(b), reveals the
inappropriateness of the use of AAFT.
The null hypothesis is rejected even for $i\!=\!2,\ldots,11$, i.e., when a
linear statistic is used.  
On the other hand, using IAAFT-1 or IAAFT-2 surrogate data only the
$\rho(1,i)$ for $i \geq m+2$, i.e., involving nonlinear terms, give
$S$ over the $2\sigma$ level. 
For only linear terms, $S$ is at the $2\sigma$ level using IAAFT-1
surrogate data and falls to zero when IAAFT-2 surrogate data are used instead.
Employing as discriminating statistic the difference of $\rho(1)$
after, for example, the inclusion of the $x_i^2$ term, i.e., $q \!=\!
\rho(1,12) - \rho(1,11)$, gives for AAFT $S \!=\! 1.76\sigma$, for IAAFT-1
$S \!=\! 3.35\sigma$, and for IAAFT-2 $S \!=\! 4.05\sigma$.       

So, even for short time series, for which IAAFT-1 cannot match perfectly the
autocorrelation, it turns out that the IAAFT algorithm distinguishes correctly
nonlinear dynamics while following the AAFT algorithm one can be fooled or at
least left in doubt as to the rejection of the null hypothesis.  
Here, we had already strong evidence from previous works that nonlinear
dynamics is present, and used this example to demonstrate the confusion AAFT
method can cause. 
However, if one checks first the autocorrelation of AAFT
[Fig.\ \ref{SunspotsAutoc}(a)], then these results should be anticipated
[Fig.\ \ref{SunspotsCC}(a) and (b)].

% -----------------------
\subsection{The AE Index data} 
% -----------------------

We examine here a geophysical time series, the auroral electrojet index
(AE) data from magnetosphere \cite{Klimas96}.  
Surrogate data tests for the hypothesis of nonlinear dynamics have been
applied to records of AE index of different time scales and sizes with
contradictory results
\cite{Prichard92b,Prichard93,Pavlos94,Pavlos99,Pavlos99b}.   
Here, we use a long record of six months, but smoothly resampled to a final
data set of $4096$ samples [see Fig.\ \ref{realseries}(a)].
% ----------------------
%           Fig.5
% ----------------------
The amplitude distribution of the AE index data is non-Gaussian and has a bulk
at very small magnitudes and a long tail along large magnitudes. 
This is so, because the AE index is characterized by resting periods
interrupted by periods of high activity probably coupled to the storms of
solar wind. 

For the autocorrelation, it turns out that the AAFT algorithm gives positive
bias in this case, i.e. the AAFT surrogates are significantly more correlated
than the original.   
The $R_y(\tau)$ of the ordered noise data $\{ y \}$ are slightly larger than
$R_x(\tau)$, which, according to Section\ \ref{Algorithms}, is a sign that
under the null hypothesis the $h$ transform is not monotonic.  
Also $R_z(\tau) \! \leq \! |R_y(\tau)|$ holds so that $h$ seems to have been
simulated successfully. 
On the other hand, the IAAFT-1 surrogates match almost perfectly the
autocorrelation and represent exactly the null hypothesis (therefore IAAFT-2
surrogate data are not used here).

The $\rho(1,i)$ from the Volterra polynomial fit on the original AE index
shows a characteristic improvement of fitting with the addition of the first
nonlinear term (see Fig.\ \ref{AE8CC}). 
% ----------------------
%           Fig.6
% ----------------------
This result alone gives evidence for nonlinear dynamics in the AE index.
However, the surrogate data test using the AAFT does not support this finding
due to the bias and variance in the autocorrelation.
To the contrary, as shown in Fig.\ \ref{AE8CC}(b), it gives the confusing
pattern that the null hypothesis is marginally rejected at the $95\%$
confidence level with linear discriminating statistics [$\rho(1,i)$ for
$i\!=\!2,\ldots,11$], but not rejected with nonlinear statistics
[$\rho(1,i)$ for $i\!=\!12,\ldots,66$].
The IAAFT algorithm is obviously proper here.
The $\rho(1,i)$ for the IAAFT-1 follows closely the $\rho(1,i)$ for the
original only for the linear fitting [Fig.\ \ref{AE8CC}(a)]. 
Consequently, the significance changes dramatically with the inclusion of the
first nonlinear term from $1\sigma$ to $7\sigma$ and stabilizes at this level
for all $i\!=\!12,\ldots,66$ [Fig.\ \ref{AE8CC}(b)].

The discrimination of the original AE data from AAFT surrogates can be still
achieved by employing the discriminating statistic $q \!=\! \rho(1,12) -
\rho(1,11)$ giving $S \!=\! 5.3\sigma$ (and $S \!=\! 6.4\sigma$ for IAAFT-1).  
Actually, the nonlinearity indicated from the Volterra polynomial fit is
very weak and can be marginally detected with other discriminating statistics
\cite{Pavlos99,Pavlos99b}. 
For example, a local fit would give the erroneous result that the null
hypothesis is marginally rejected using AAFT, but not using IAAFT, because the
local fit is mainly determined by the linear correlations.  
In particular, a fit with LAM ($m\!=\!10$, $\tau\!=\!1$, $k\!=\!10$) gave
for $\rho(1)$ significance $S \!=\! 1.84\sigma$ for AAFT and only $S \!=\!
0.3\sigma$ for IAAFT. 

% -----------------------
\subsection{Breath rate data} 
% -----------------------

The next real data set is from the breath rate of a patient with sleep apnea.
The time series consists of the first $4096$ samples of the set B of the Santa
Fe Institute time series contest \cite{Rigney94} [see Fig.\
\ref{realseries}(b)].   
This time series is characterized by periods of relaxation succeeded by periods
of strong oscillations and follows a rather symmetric amplitude distribution
but not Gaussian (more spiky at the bulk). 
These data are also believed to be nonlinear, but it is not clear whether the
nonlinearity is autonomous or merely due to nonlinear coupling with the heart
rate \cite{Schreiber98b}.   

The breath rate time series does not have strong linear correlations.
However, AAFT gives again bias in the autocorrelation, but not large variance 
while IAAFT-1 matches perfectly the original autocorrelation (therefore
IAAFT-2 is not used here).   

The Volterra polynomial fit, shown in Fig.\ \ref{B11CC}, reflects exactly
the results on the autocorrelation.    
% ----------------------
%           Fig.7
% ----------------------
For the linear terms, the $\rho(1,i)$ for AAFT are rather concentrated at a
level roughly $10\%$ lower than the $\rho(1,i)$ for the original data. 
This large difference combined with the small variance does not validate the
comparison of AAFT surrogate data to the original data with any nonlinear
tool sensitive to data correlations.
For the IAAFT-1, the situation is completely different. 
The perfect matching in $\rho(1,i)$ for the linear terms in combination with
the abrupt rise of the $\rho(1,i)$ of the breath rate data after the inclusion
of the second (not first!) nonlinear term, constitutes a very confident
rejection of the null hypothesis at the level of at least $25\sigma$, as shown
in Fig.\ \ref{B11CC}(b).      
 
It seems that for the modeling of the breath rate data, the interaction of
$x_i$ and $x_{i-1}$ (term $13$) is very important. Using the discriminating
statistic $q \!=\! \rho(1,12) - \rho(1,11)$ as before gives significance
around $3\sigma$ for both AAFT and IAAFT, but using $q \!=\! \rho(1,13) -
\rho(1,12)$ instead gives significance about $80\sigma$ for both AAFT and
IAAFT.   

% -------------------
\subsection{EEG data} 
% -------------------

The last example is the measurement of $4096$ samples from an arbitrary
channel of an EEG recording assumed to be during normal activity [actually,
the record is from an epileptic patient long before the seizure, see Fig.\
\ref{realseries}(c)].  
Though the deviation of the amplitude distribution from Gaussian is small, the
AAFT algorithm gives again large bias in the autocorrelation, while IAAFT-1
achieves good matching.   
Particularly, the condition $R_z(\tau) \! \leq \! |R_y(\tau)|$ does not hold
here for small $R$ values implying that the bias in autocorrelation may also
be due to unsuccessful simulation of $h$ (in the step\ 3 of the AAFT
algorithm). 

This EEG time series does not exhibit nonlinearity, at least as observed from
the one step ahead fit with Volterra polynomials (Fig.\ \ref{EEGCC}).  
% ----------------------
%           Fig.8
% ----------------------
Obviously, the difference in $\rho(1,i)$ between original and AAFT surrogates
wrongly suggests rejection of the null hypothesis when the nonlinear
Volterra polynomial fit (terms $>11$) is used as discriminating statistic.  
This is solely due to the bias in autocorrelation as this difference remains
also for the linear terms. 
For IAAFT-1, there is a small difference in $\rho(1,i)$ for the linear terms,
as shown in Fig.\ \ref{EEGCC}(a), though IAAFT-1 seems to give good matching
in the autocorrelation.   
This probably implies that the $\rho$ from the linear fit amplifies even small
discrepancies in autocorrelation, not detected by eye-ball judgement. 
Moreover, this small difference in $\rho(1,i)$ is significant, as shown in
Fig.\ \ref{EEGCC}(b), because again IAAFT-1 tends to give dense statistics.  
Remarkably, the significance degrades to less than $2\sigma$ when
nonlinear terms are added.

We employ IAAFT-2 surrogate data as well [see Fig.\ \ref{EEGCC}(a)]. 
These do not match perfectly the original amplitude distribution (especially
at the bulk of the distribution), but possess exactly the same linear
correlations as the original, as approved also from the linear fit in Fig.\
\ref{EEGCC}(a). 
For the IAAFT-2 surrogates, the significance from the $\rho(1,i)$ is correctly
less than $2\sigma$ for both the linear and nonlinear terms of the polynomial
fit, as shown in Fig.\ \ref{EEGCC}(b). 

We want to stress here that the results on the EEG data are by no means
conclusive, as they are derived from a simulation with a single tool
(polynomial fit) on a single EEG time series. 
However, they do insinuate that the use of AAFT surrogate data in the numerous
applications with EEG data should be treated with caution at least when a
striking difference between the original data and the AAFT surrogate data
cannot be established, which otherwise would rule out that the difference is
solely due to biased autocorrelation.  
 
%====================================================
\section{Discussion}
%====================================================

The study on the two methods for the generation of surrogate data that
represent the null hypothesis of Gaussian 
correlated noise undergoing nonlinear static distortion revealed
interesting characteristics and drawbacks of their performance.
The most prominent of the two methods, the amplitude adjusted Fourier
transform (AAFT) surrogates, can represent successfully the null hypothesis
only if the static transform $h$ is monotonic. 
This is an important generic characteristic of the AAFT algorithm and not just
a technical detail of minor importance as treated in all applications with
AAFT so far
\cite{Prichard93,Rombouts95,Yip95,Pritchard95,Pradhan96,Ivanov96,Theiler96,Menendez97,Stam97,Govindan98}.
The bias in autocorrelation induced by the non-monotonicity of $h$ can lead
to false rejections of the null hypothesis. 

Our simulations revealed a drawback for the other method, the iterated AAFT
(called IAAFT here), which was not initially expected.  
Depending on the data type, the iterative algorithm may naturally terminate
while the matching in autocorrelation is not yet exact (we call the derived
surrogate data IAAFT-1). 
In this case, all IAAFT-1 surrogate data achieve approximately the same level
of accuracy in autocorrelation. 
Thus, the variance of autocorrelation is very small and therefore the
mismatching becomes significant. 
Consequently, applying a nonlinear statistic sensitive to data correlations
gives also significant discrimination. 
So, when using IAAFT surrogate data, the exact matching in autocorrelation
must be first guaranteed and then differences due to nonlinearity become more
distinct due to the small variance of the statistics on IAAFT.
In cases the IAAFT-1 data set does not match exactly the original
autocorrelation, we suggest to use a second data set derived from the same
algorithm, IAAFT-2, which possesses exactly the same linear correlations as
the original and may slightly differ in the amplitude distribution. 
Our simulations suggest the use of IAAFT-2 in general, but there may
be cases where a detailed feature of the amplitude distribution should be
preserved (e.g. data outliers of special importance) and then IAAFT-1 should
be used instead. 

The application of the AAFT and IAAFT algorithms to real world data
demonstrated the inappropriateness of AAFT and the ''too good'' significance
obtained with IAAFT surrogates if nonlinearity is actually present. 
The results generally suggest that one has first to assure a good matching in
autocorrelation of the surrogate data to the original before using them
further to compute nonlinear discriminating statistics and test the null
hypothesis.  
If a bias in autocorrelation is detected, statistical difference in the
nonlinear statistics may also occur and then the rejection of the null
hypothesis is not justified by a high significance level
because it can be just an artifact of the bias in autocorrelation.

One can argue that when $h$ is not invertible then the assumption that the
examined time series stems from a Gaussian process cannot be assessed by this
test because there is not one to one correspondence between the measured time
series $\{ x \}$ and the Gaussian time series $\{ s \}$, where $x = h(s)$. 
We would like to stress that the hypothesis does not yield a single Gaussian
process, but any Gaussian process that under $h$ (even not monotonic) {\em can
give} $\{ x \}$, i.e., the existence of multiple solutions is not excluded.
More precisely, the null hypothesis states that the examined time series
belongs to the family of statically transformed Gaussian data with linear
properties and deviation from the Gaussian distribution determined by the
corresponding sample quantities of the original time series.
Thus the surrogate data generated under the two conditions (matching in
autocorrelation and amplitude distribution) may as well be considered as
realizations of different Gaussian processes statically transformed under $h$. 
Differences within the different underlying linear processes are irrelevant
when the presence of nonlinearity is investigated.

Concerning the discriminating statistics, our findings with synthetic and real
data suggest that local models, such as the local average map, are not always 
suitable to test the null hypothesis and can give confusing results. 
On the other hand, the Volterra polynomial fit turned out to be a useful
diagnostic tool for detecting dynamic nonlinearity directly on the original
data as well as 
verifying the performance of the surrogate data because it offers direct
detection of changes in the discriminating statistic from the linear to
nonlinear case.   

%====================================================
\section*{Acknowledgements}
%====================================================

The author thanks Giorgos Pavlos for providing the AE index data, P{\aa}l
Larsson for providing the EEG data, and Holger Kantz for his valuable comments
on the manuscript. 

%====================================================
% And the bibliography, Using BibTeX

%====================================================

\clearpage

\widetext

%====================================================
% FIGURES 
%====================================================

%====================================================
\begin{figure}[htb] % 1
\caption{Autocorrelation for lag $1$ as a function of the power exponent of
the transform $h$ for an AR(1) process and for AAFT surrogates in (a) and
IAAFT-1 surrogates in (b).
The average $R(1)$ for $\{ s \}$, $\{ x \}$, $\{ y \}$ and $\{ z \}$ are as
denoted in the legend. 
The $R_s(1)$ for $100$ realizations of $\{ s \}$ are shown with the short line
segments on the right upper side of the figures. 
For $R_x(1)$, the SD over the 100 realizations is denoted with error bars
around the average. 
For $R_y(1)$ and $R_z(1)$, the zones around the average denote the average
over $100$ realizations of the SD computed from $40$ surrogate data at each
realization.} 
\label{AR1Autoc}
\end{figure}
%====================================================

%====================================================
\begin{figure}[htb] % 2
\caption{Rejections of the null hypothesis using two discriminating statistics
plotted as a function of the power exponent of the transform $h$ for the data
used in Fig.\ \ref{AR1Autoc}.   
In (a), the statistic is the correlation coefficient of the one step ahead
fit $\rho(1)$ using an AR(1) model and in (b) the $\rho(1)$ using a local
average mapping (LAM).  
The rejections yield AAFT, IAAFT-1 and IAAFT-2 as denoted in the legend.}   
\label{AR1Rej}
\end{figure}
%====================================================

%====================================================
\begin{figure}[htb] % 3
\caption{Autocorrelation $R(\tau)$ for the sunspot numbers (thick black line)
and its $40$ surrogates (gray thin lines), AAFT in (a), and IAAFT-1 in (b). 
In (a), $R(\tau)$ for the reordered noise ($\{ y \}$ data) is also shown,
with a thin black line for the average value and error bars for the SD.} 
\label{SunspotsAutoc}
\end{figure}
%====================================================

%====================================================
\begin{figure}[htb] % 4
\caption{(a) Correlation coefficient $\rho(1,i)$ as a function of the terms of
the Volterra polynomials ($m=10$, $d\!=\!2$, $i\!=\!1,\ldots,66$) for the
sunspot numbers (thick black line) and its $40$ AAFT, IAAFT-1 and IAAFT-2
surrogates (gray thin lines) in the three plots as indicated.  
(b) Significance $S$ of $\rho(1,i)$ for AAFT, IAAFT-1 and IAAFT-2.  
The vertical line is to stress the insertion of the first nonlinear term.} 
\label{SunspotsCC}
\end{figure}
%====================================================

%====================================================
\begin{figure}[htb] % 5
\caption{Three real time series of $4096$ samples each. 
(a) The AE index time series measured every minute over the second half year
of 1978 and smoothed to a time resolution of $\tau_s = 64$min; the data range
is $[6,1500]$.
(b) The breath rate time series sampled at a sampling time $\tau_s = 0.5$sec;
the data range is $[-12489,32740]$.
(c) The EEG time series sampled at a sampling time $\tau_s = 0.005$sec; the
data range is $[1797,2152]$.}
\label{realseries}
\end{figure}
%====================================================

%====================================================
\begin{figure}[htb] % 6
\caption{(a) Correlation coefficient $\rho(1,i)$ as a function of the terms of
the Volterra polynomials ($m=10$, $d\!=\!2$, $i\!=\!1,\ldots,66$) for the
AE index (thick black line) and its $40$ AAFT and IAAFT-1 surrogates (gray
thin lines) in the two plots as indicated.    
(b) Significance $S$ of $\rho(1,i)$ for AAFT and IAAFT-1.  
The vertical line is to stress the insertion of the first nonlinear term.} 
\label{AE8CC}
\end{figure}
%====================================================

%====================================================
\begin{figure}[htb] % 7
\caption{(a) Correlation coefficient $\rho(1,i)$ as a function of the terms of
the Volterra polynomials ($m=10$, $d\!=\!2$, $i\!=\!1,\ldots,66$) for the
breath rate data set (thick black line) and its $40$ AAFT and IAAFT-1
surrogates (gray thin lines) in the two plots as indicated.    
(b) Significance $S$ of $\rho(1,i)$ for AAFT and IAAFT-1.  
The vertical line is to stress the insertion of the first nonlinear term.} 
\label{B11CC}
\end{figure}
%====================================================

%====================================================
\begin{figure}[htb] % 8
\caption{(a) Correlation coefficient $\rho(1,i)$ as a function of the terms of
the Volterra polynomials ($m=10$, $d\!=\!2$, $i\!=\!1,\ldots,66$) for the
EEG data (thick black line) and its $40$ AAFT, IAAFT-1 and IAAFT-2
surrogates (gray thin lines) in the three plots as indicated.  
(b) Significance $S$ of $\rho(1,i)$ for AAFT, IAAFT-1 and IAAFT-2.  
The vertical line is to stress the insertion of the first nonlinear term.} 
\label{EEGCC}
\end{figure}
%====================================================

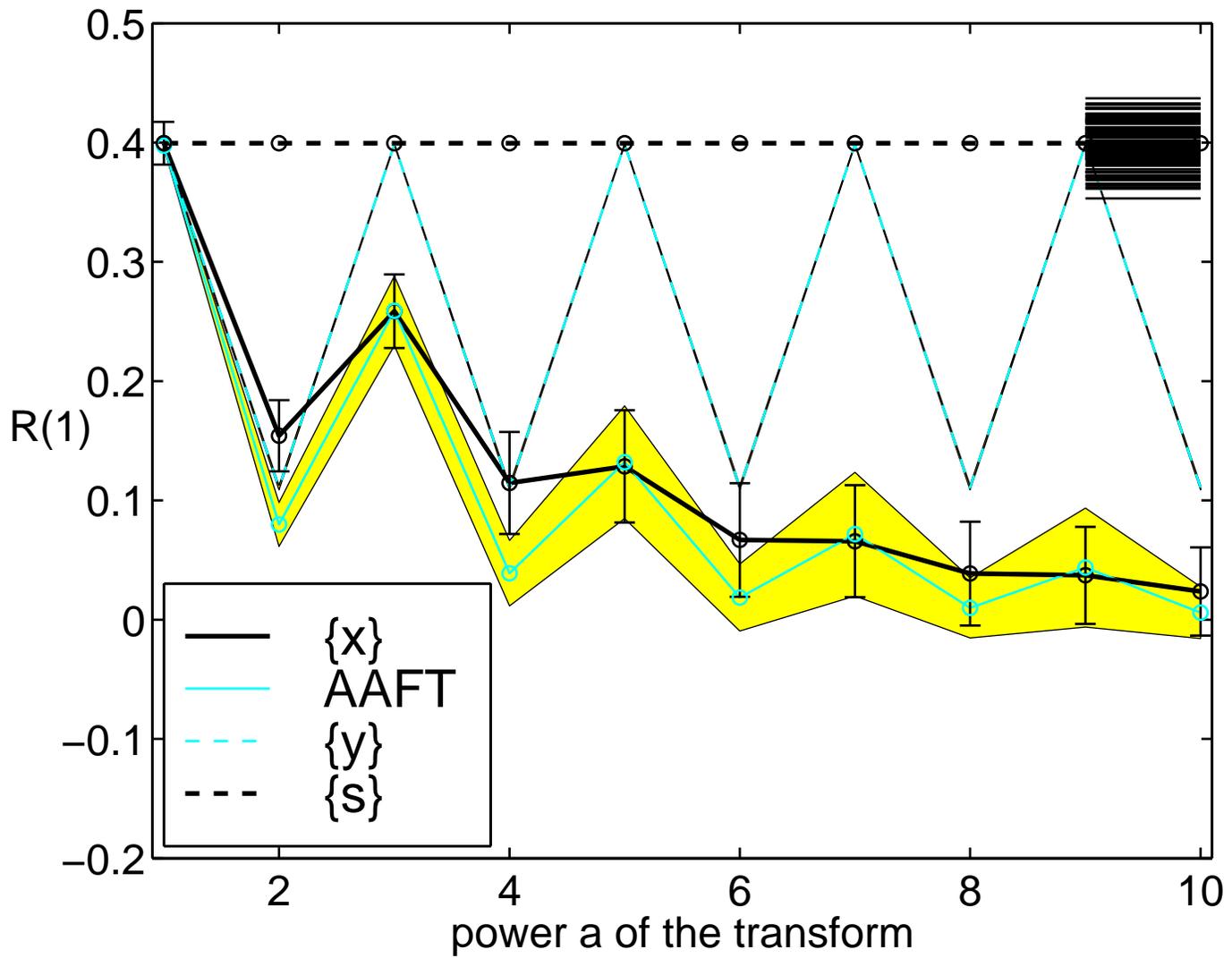
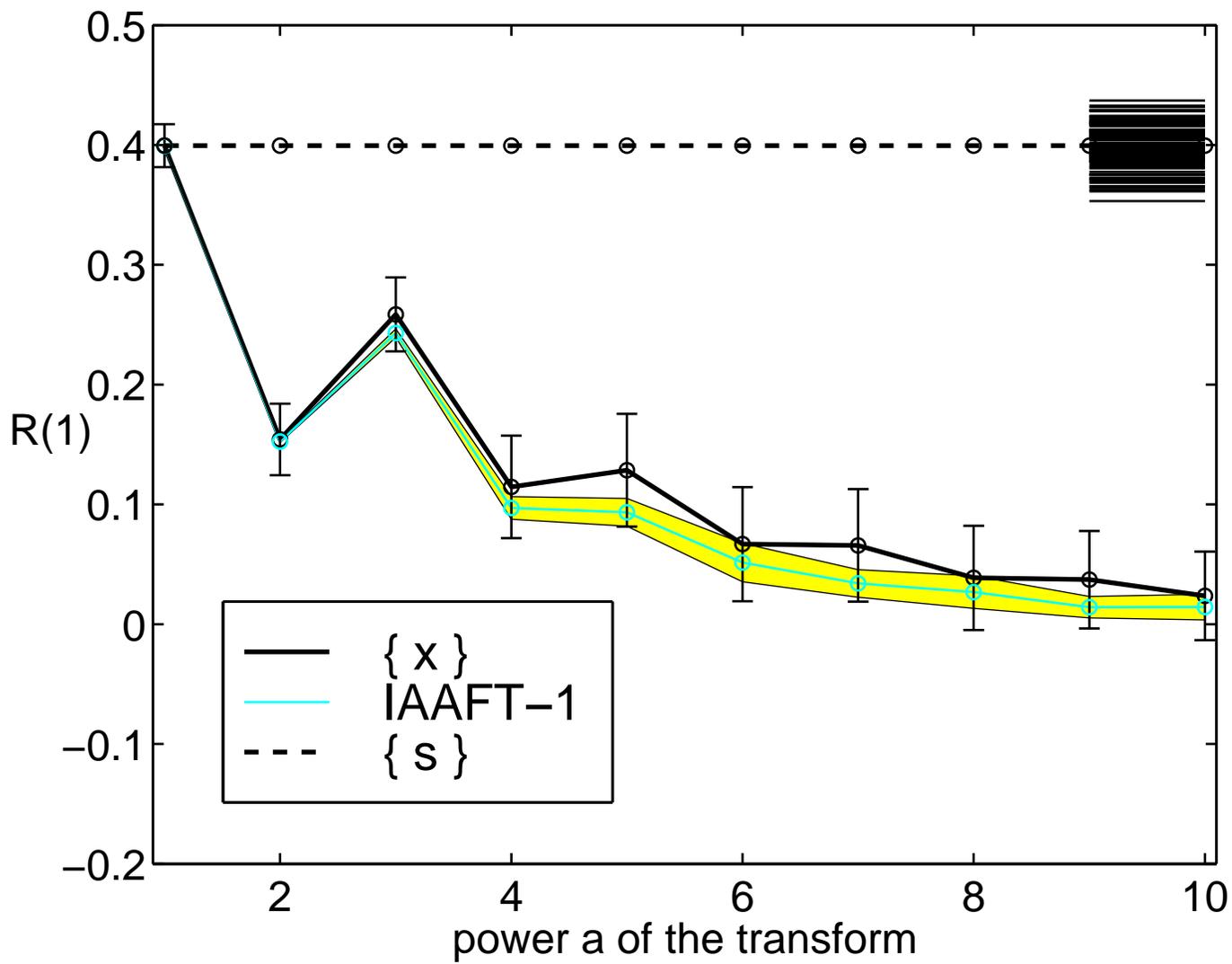
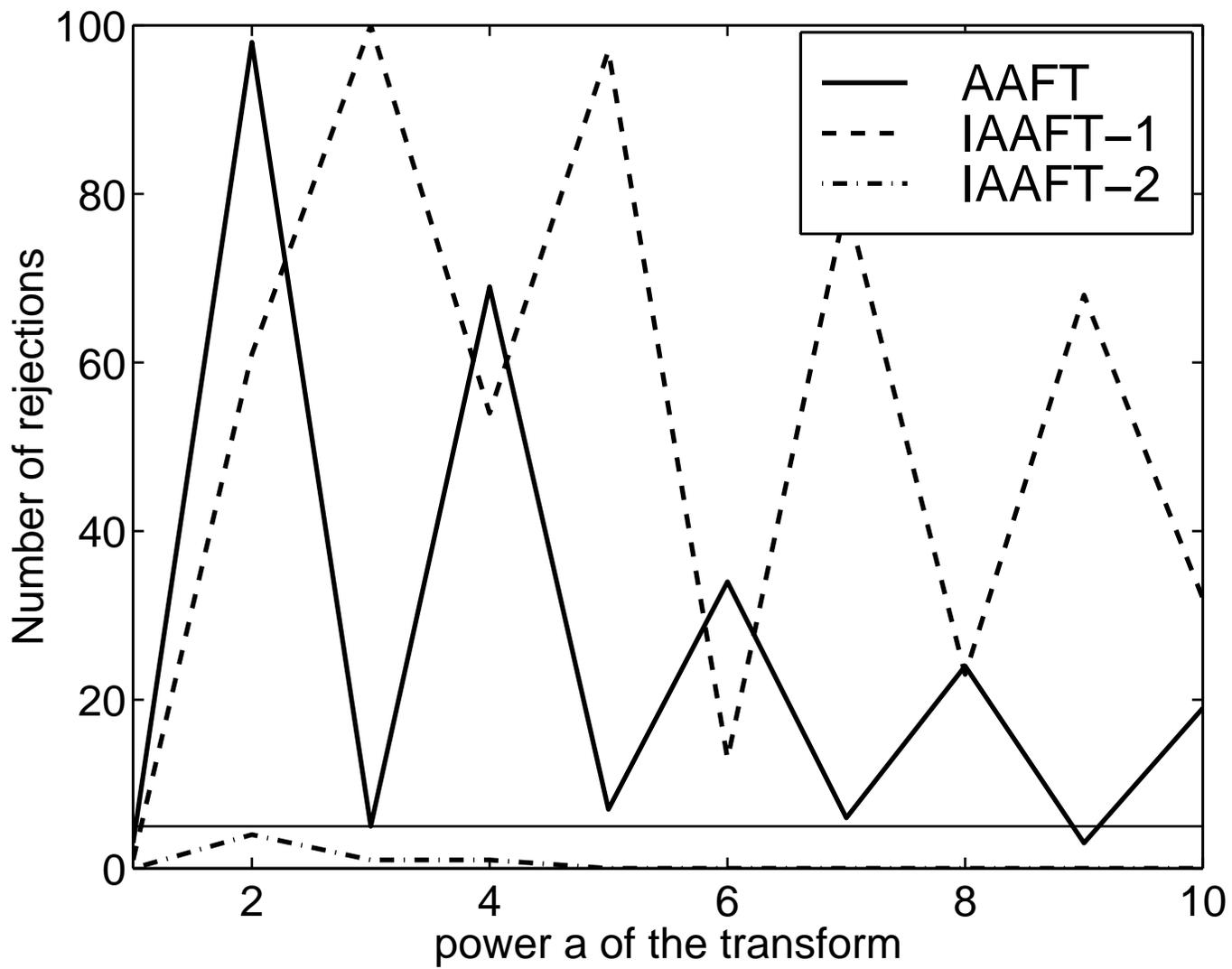
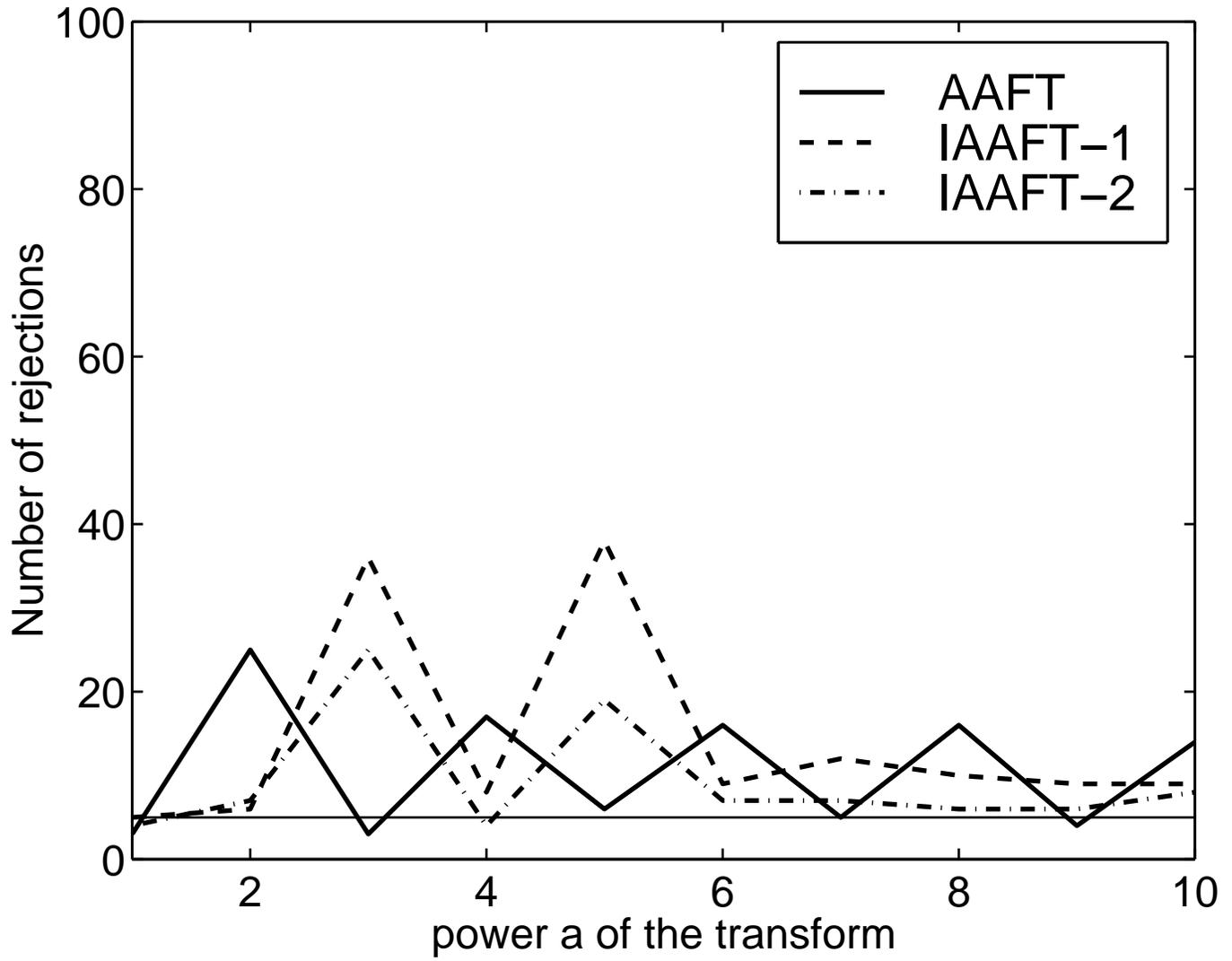
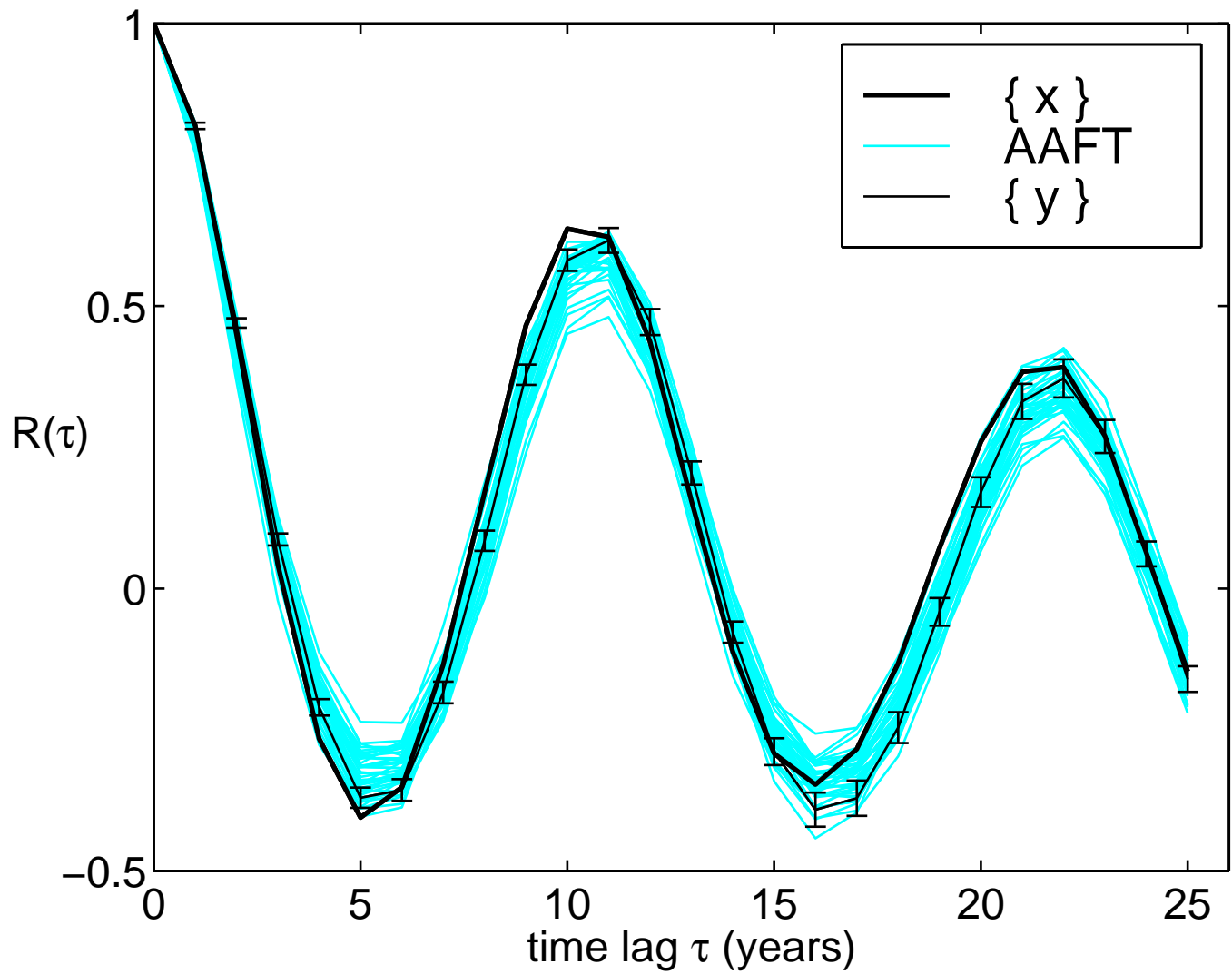
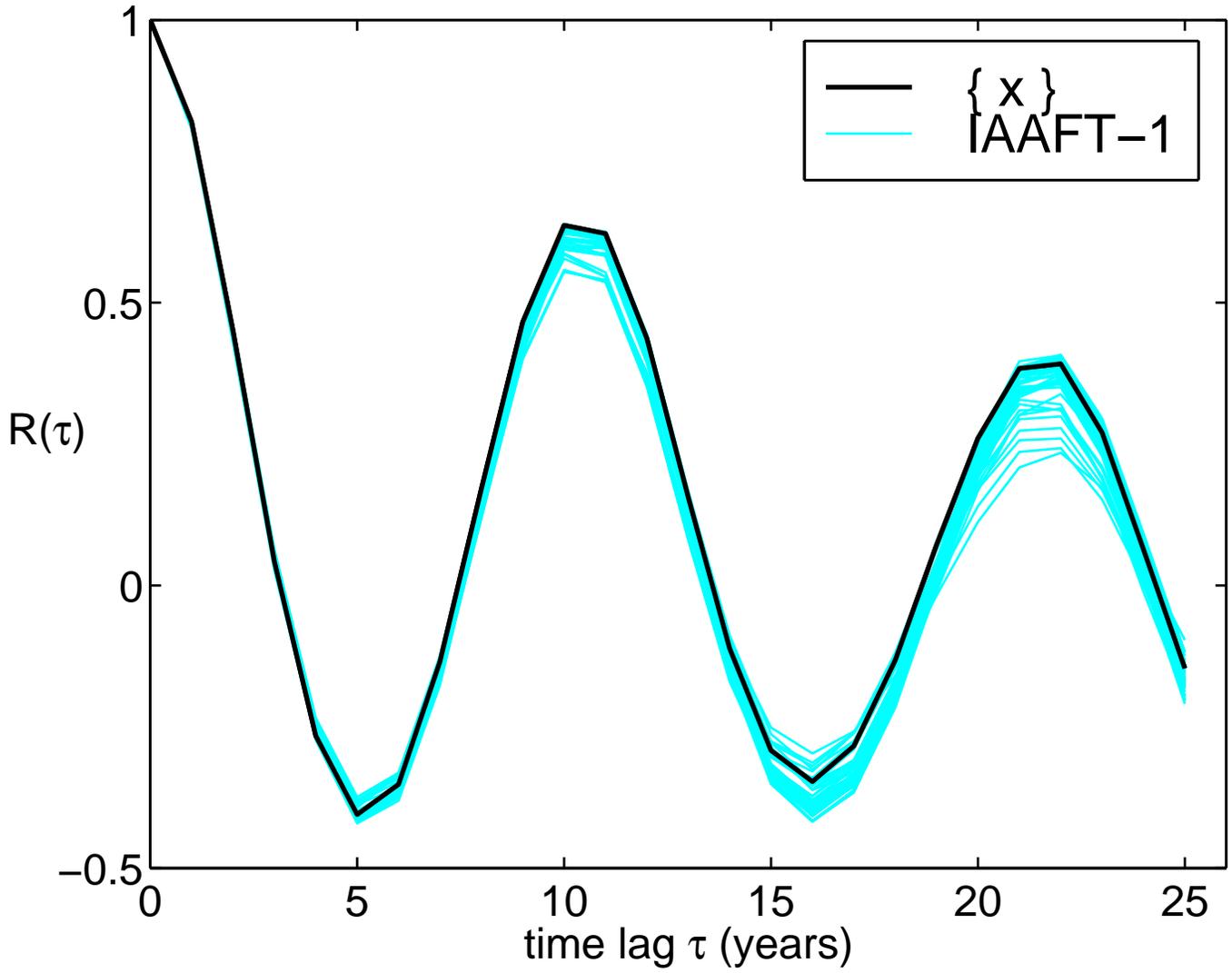
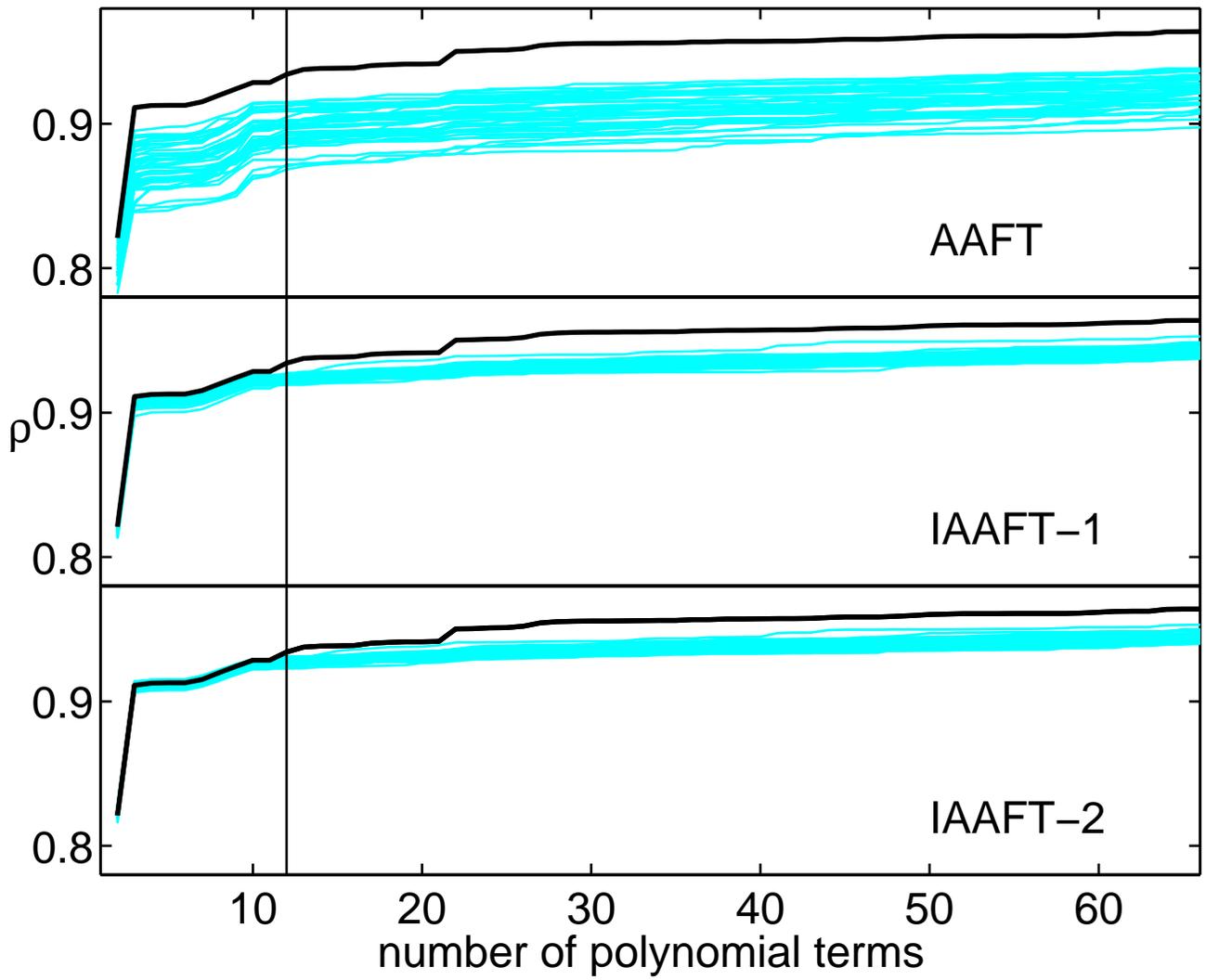
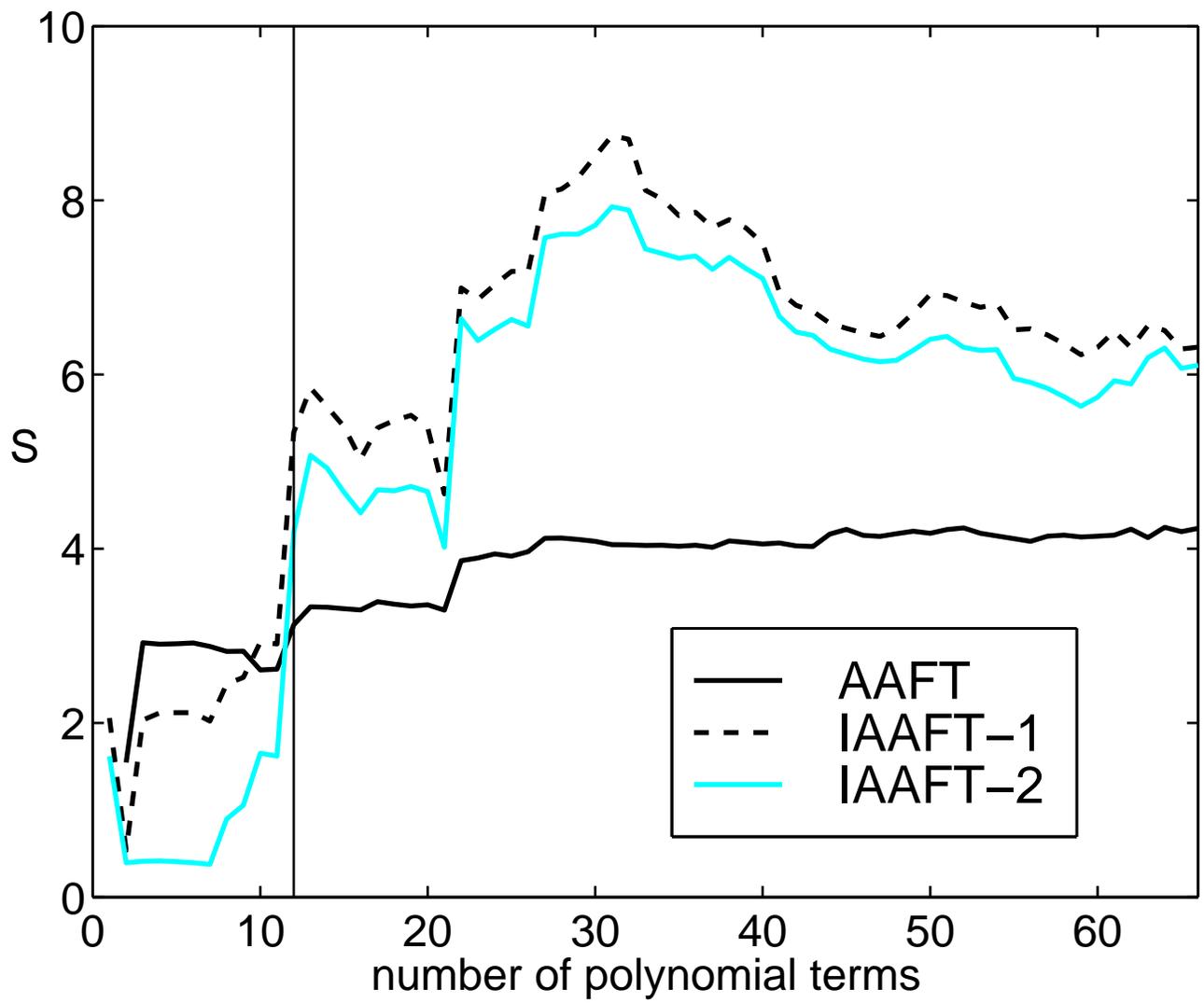
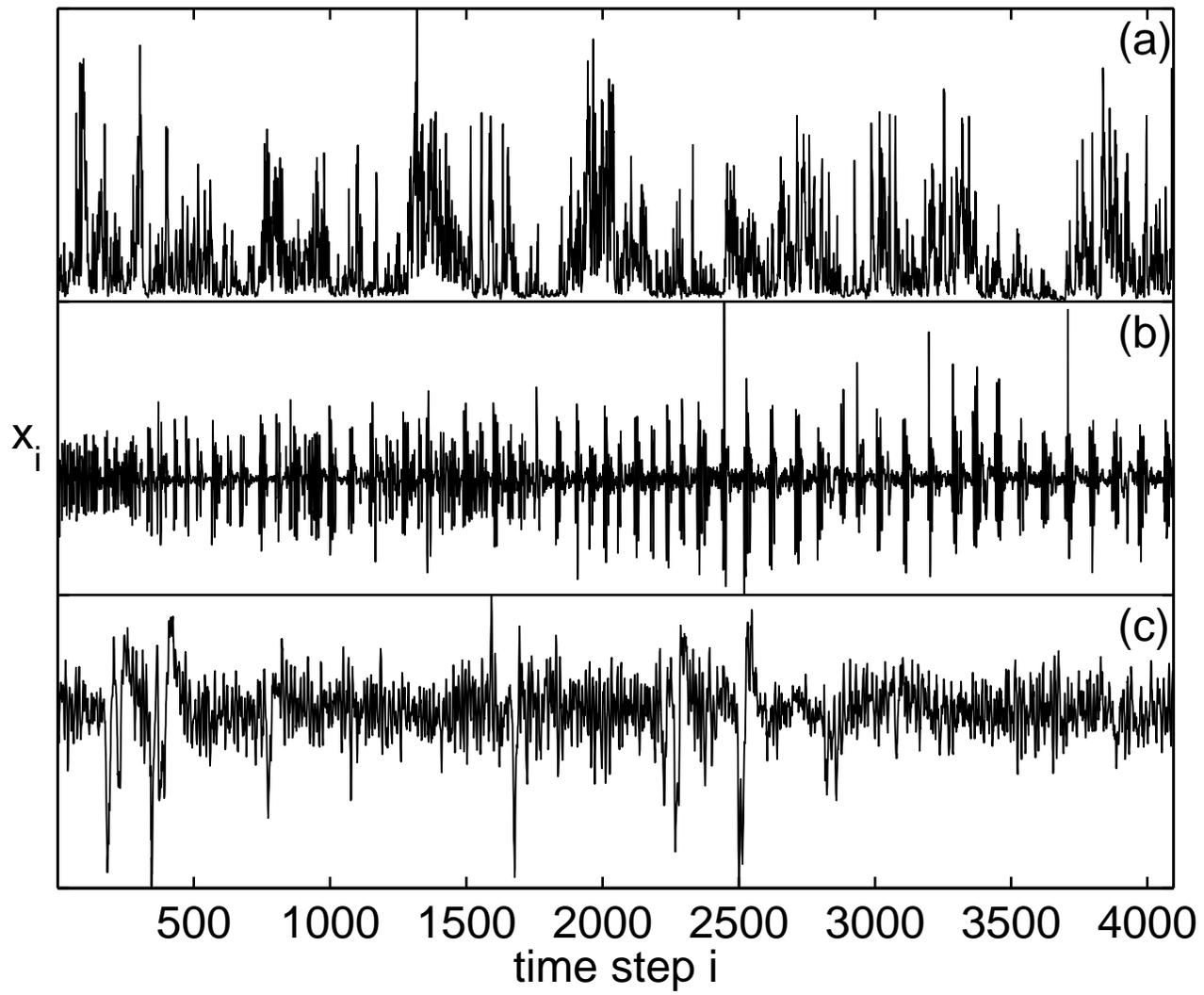
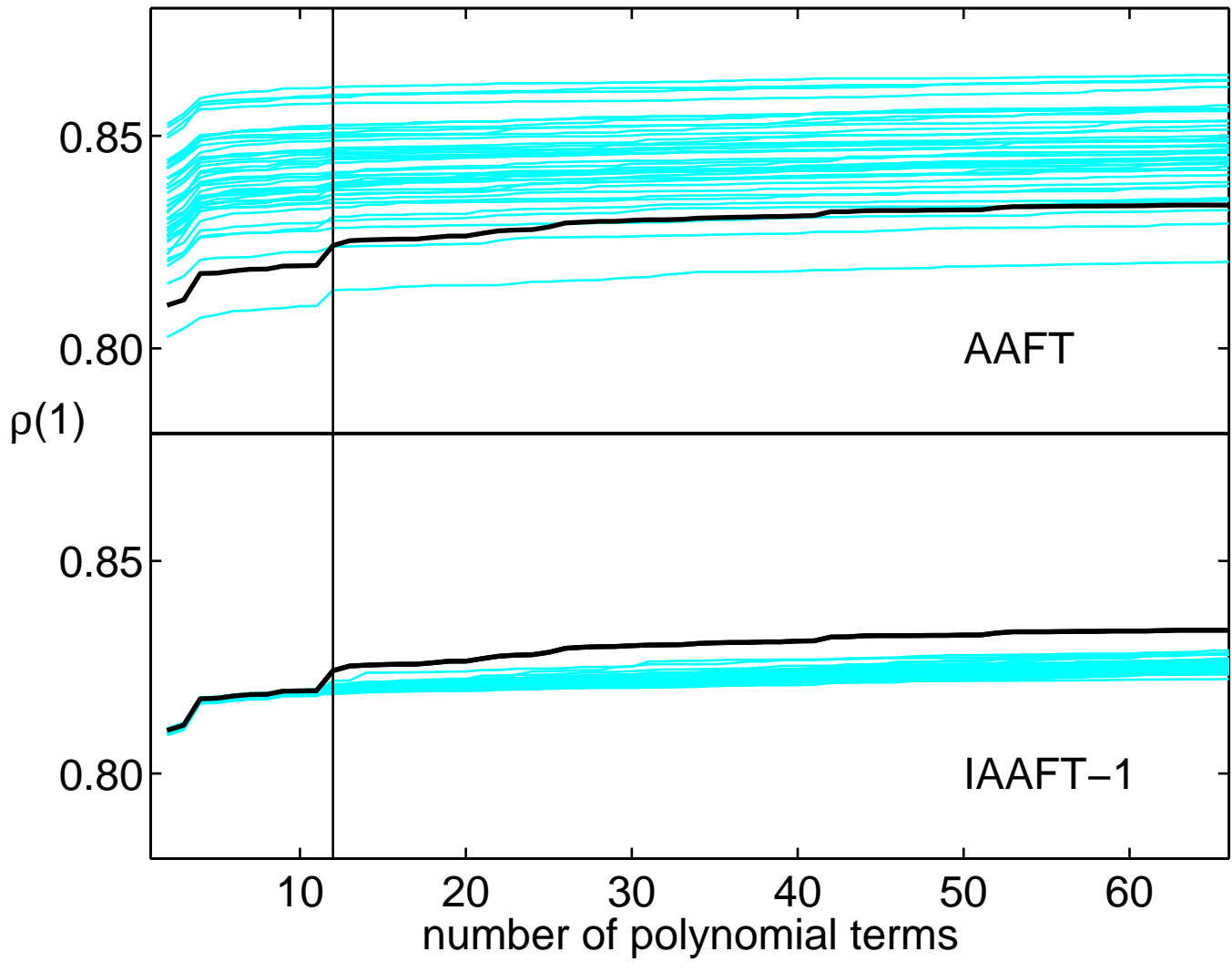
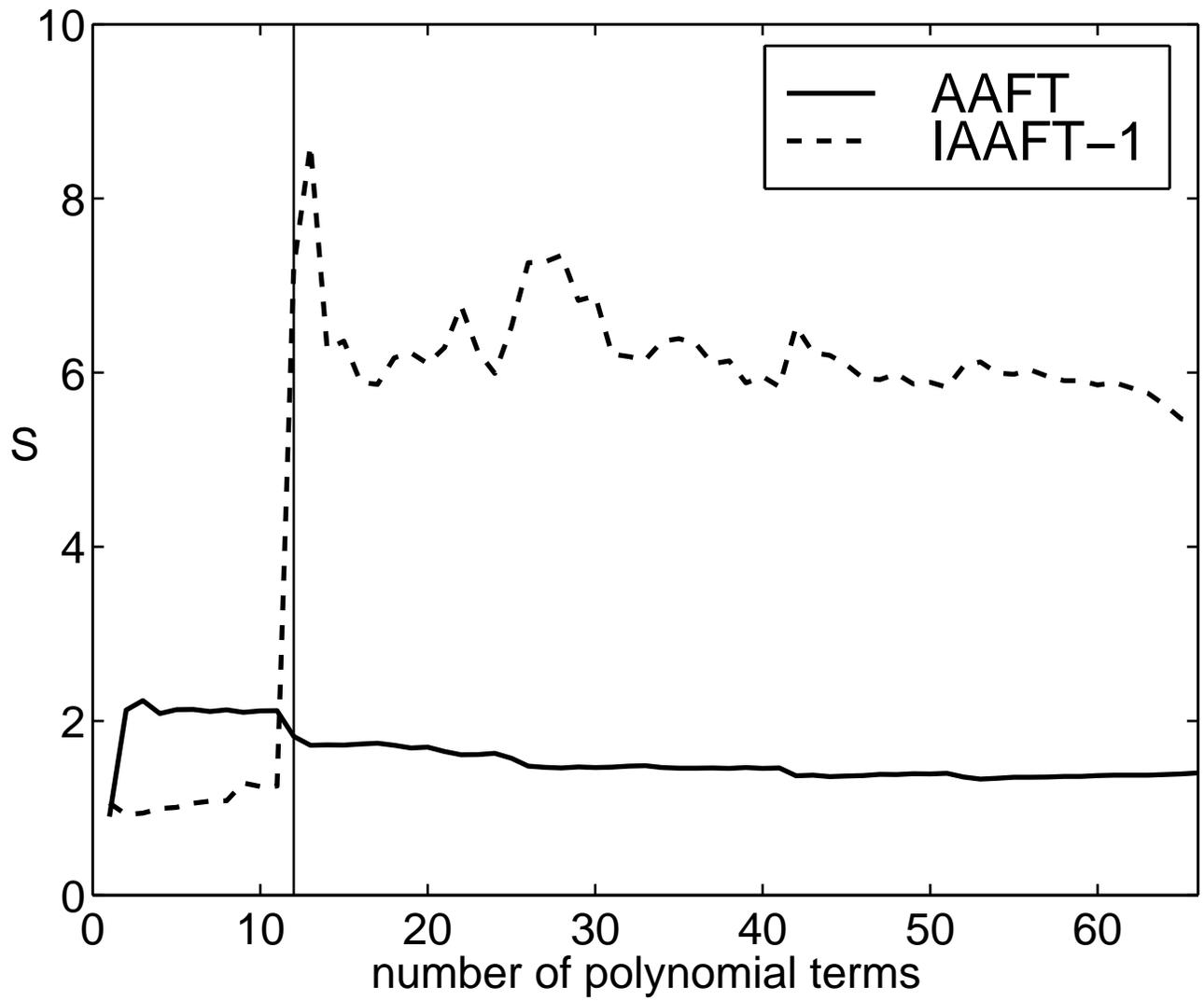
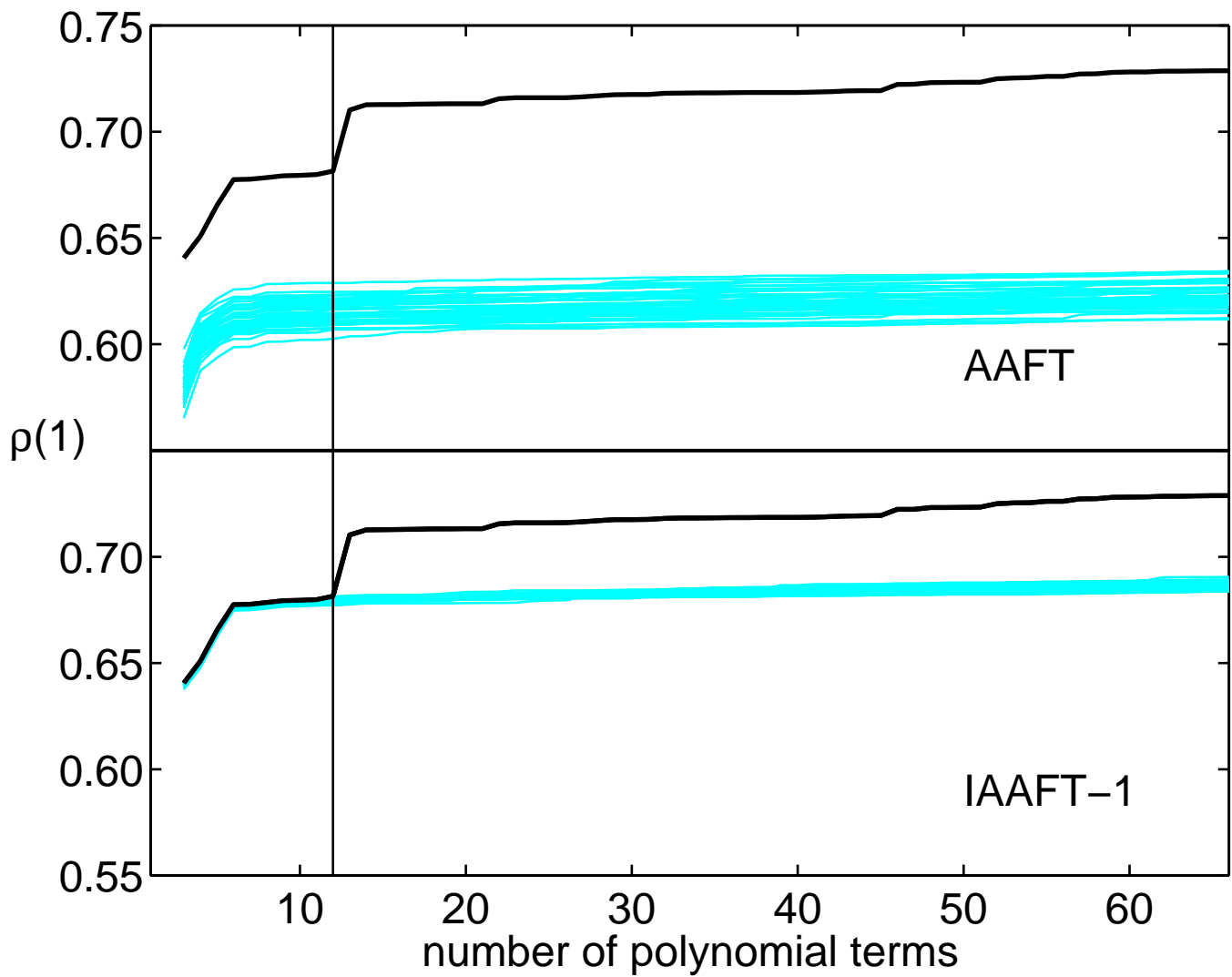
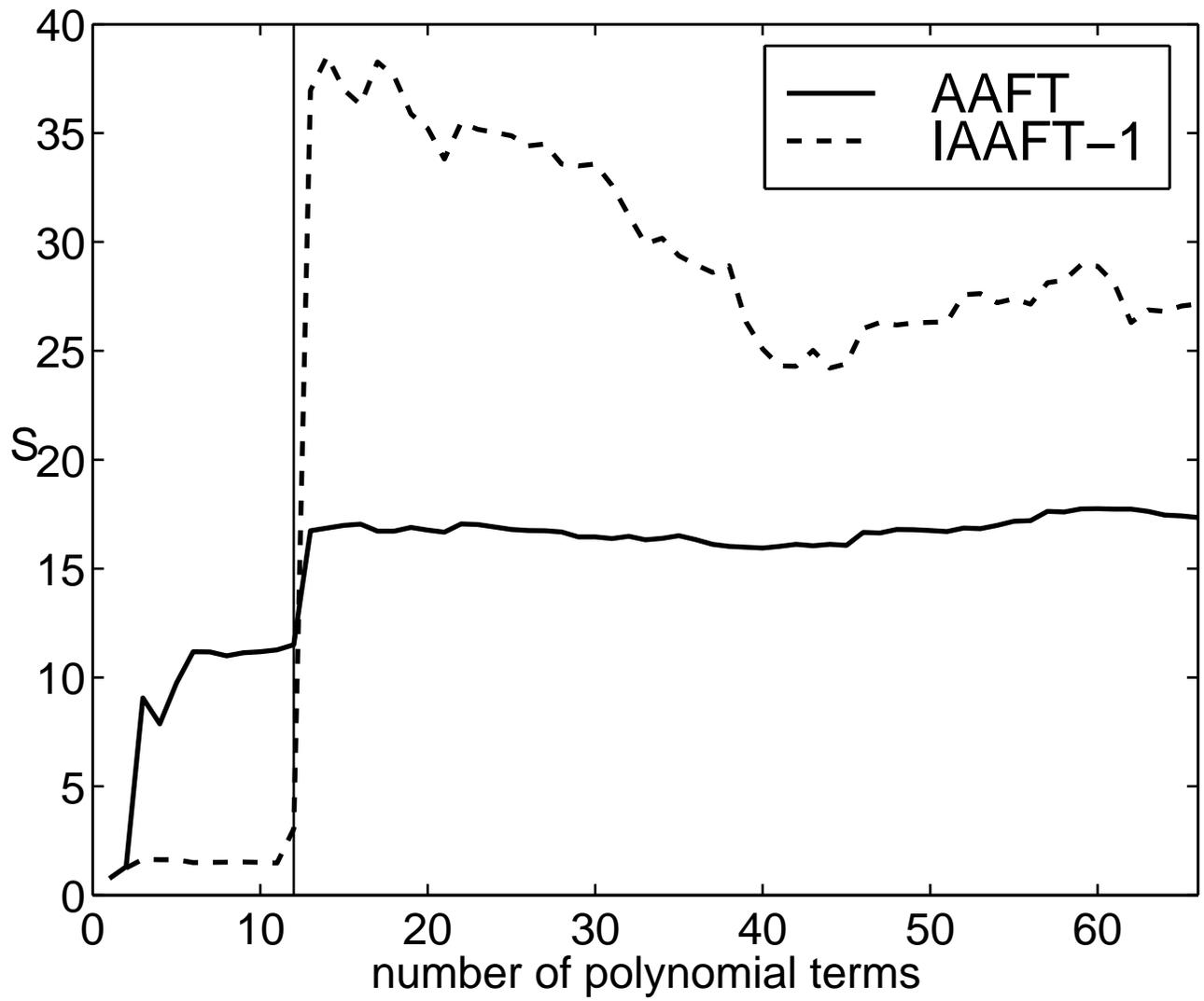
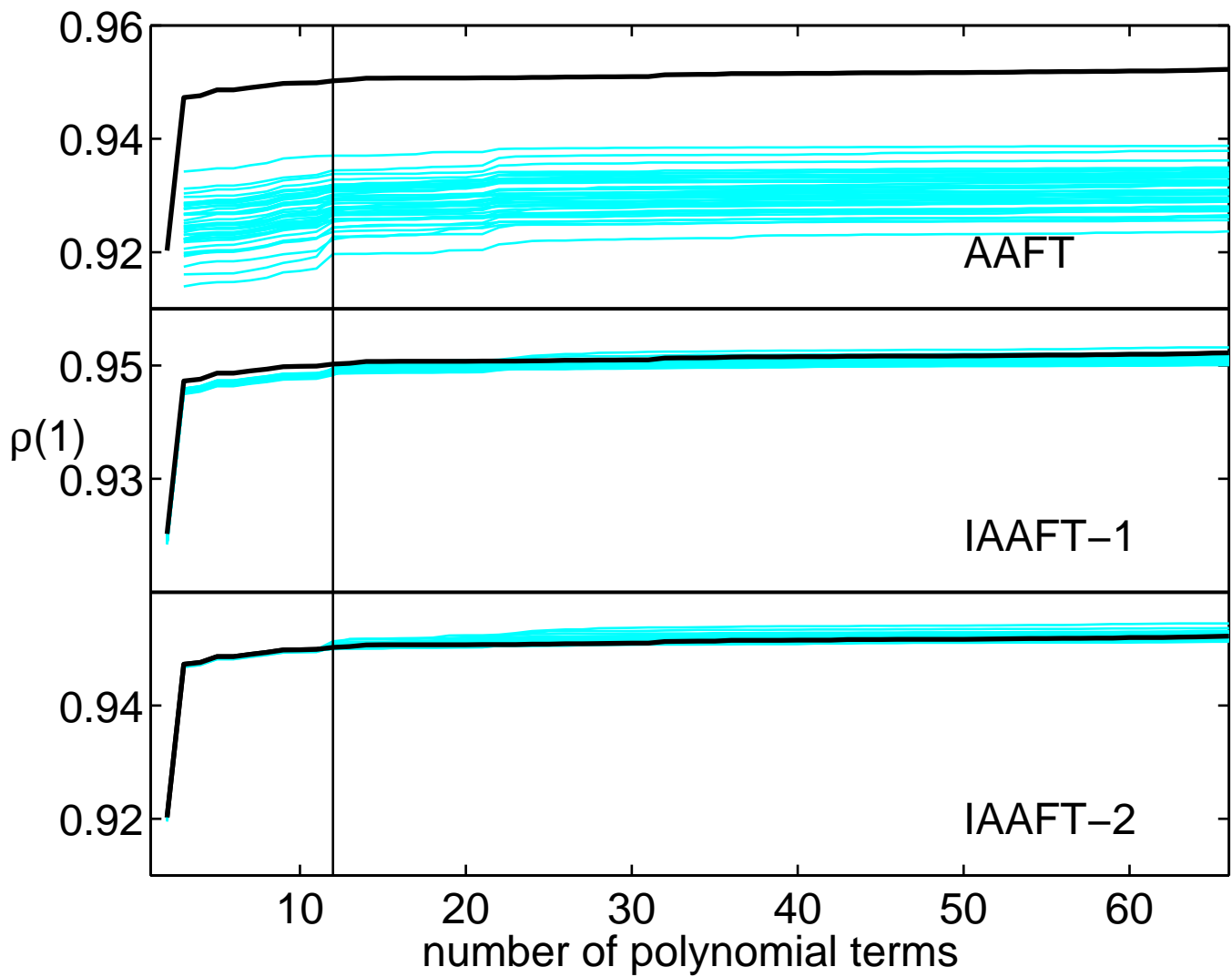
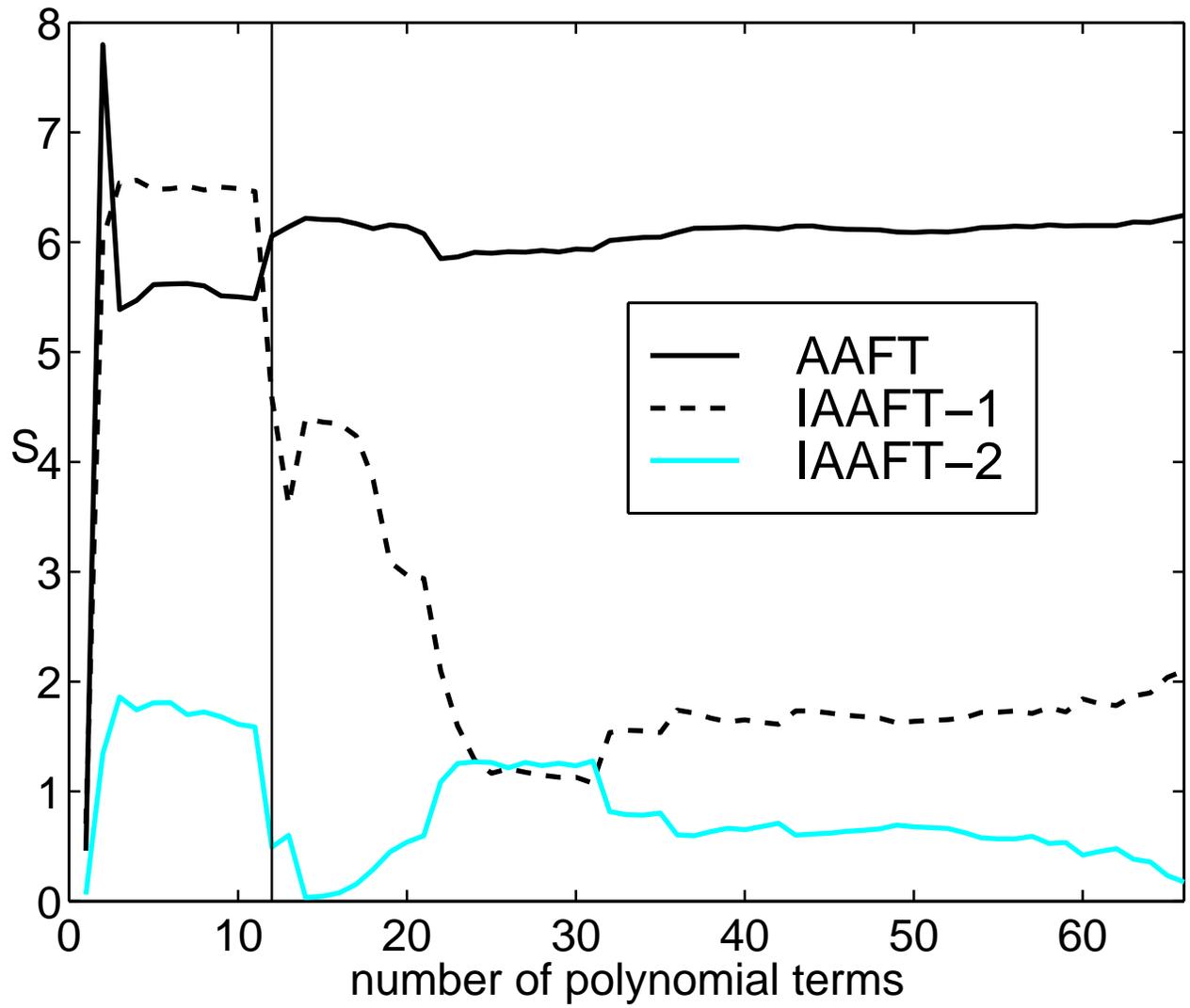
\end{document}